\documentclass[lettersize,journal]{IEEEtran}
\usepackage{amsmath,amsfonts}
\usepackage{algorithmic}
\usepackage{array}
\usepackage[caption=false,font=normalsize,labelfont=sf,textfont=sf]{subfig}
\usepackage{textcomp}
\usepackage{stfloats}
\usepackage{url}
\usepackage{verbatim}
\usepackage{graphicx}
\usepackage{xcolor}
\def\BibTeX{{\rm B\kern-.05em{\sc i\kern-.025em b}\kern-.08em
    T\kern-.1667em\lower.7ex\hbox{E}\kern-.125emX}}
\usepackage{balance}

\begin{document}
\title{
Microscopic Modeling of Surface Roughness Scattering \\
in Inversion Layers of MOSFETs Based on Ando's Linear Model
}
\author{Nobuyuki Sano 
\thanks{This work was supported in part by JSPS KAKENHI Grant Number 25K07843.
}
\thanks{The author is with Institute of Applied Physics, University of Tsukuba, Tsukuba, Ibaraki 305-8573, Japan (e-mail: sano.nobuyuki.gw@u.tsukuba.ac.jp).}}


\maketitle

{\color{black}
\begin{abstract}
Surface roughness (SR) scattering in inversion layers of bulk-MOSFETs is studied from the atomistic and quantum-mechanical viewpoints.
Contrary to the usual macroscopic landscape of the roughness deviation $\Delta \left( \mathbf{r}  \right) $, we introduce a stochastic $\Delta \left( \mathbf{r}  \right) $ {\em at each atomic site} to take account of the discontinuity of the spatial derivatives of electrostatic potential and wave-function at the semiconductor/dielectric interface, leading to an ambiguity in roughness positions.  
It is shown that SR parameters are consistent with those known from the experiments and, thus, there is no discrepancy problem associated with the roughness parameters in our model. 
The self-consistent scattering rate is derived under the framework of the Green's functions scheme: We find that the SR scattering rates are intrinsically nonlocal (non-diagonal) with respect to subband indices and greatly deviate from those based on Fermi's golden rule in the regimes of strong effective fields and/or low electron energies. As a result, the conventional SR model tends to underestimate the surface-roughness-limited mobility. 
\end{abstract}
}%

\begin{IEEEkeywords}
Surface roughness scattering, Green's function, Mobility, Scattering rate, Fermi's golden rule, Self-energy
\end{IEEEkeywords}

\section{Introduction}
\label{sec:intro}
Surface roughness (SR) scattering is known to be a dominant scattering center to limit the electron mobility in bulk-MOSFETs under the strong inversion regimes~\cite{Sun1980,Goodnick1985,Takagi1988,Takagi1994,Zhao2010}. This situation is expected to be even more true for near-future device structures such as nanoowires and nanosheets with thin-body channels. 
{\color{black}
Theoretical studies thus far are mostly based on Ando's model~\cite{Ando1977,Ando1982}, 
in which the SR transition matrix element is proportional to the deviation $\Delta \left( \mathbf{r}  \right) $ of the interface position with respect to the unperturbed semiconductor/dielectric interface. 
The model was later improved by including the dielectric screening with the random phase approximation~\cite{Jin2007TED,Jin2007}. Ando's model with screening is now considered to be the `standard' approach and widely employed to investigate the transport properties of various device structures such as  
bulk-MOSFETs~\cite{Jungemann1993,Pirovano1999}, ultrathin-body (UTB) silicon-on-insulator structures~\cite{Gamiz1999,Esseni2004,Low2005,Ishihara2006}, and UTB InGaAs-on-Insulator MOSFETs~\cite{Poljak2012}. 
In these studies, transport properties are analyzed based on the semi-classical scheme, like the Boltzmann transport equation.  
The only exception is due to Kannan and Vasileska \cite{Kannan2017}, in which the retarded Green's function was used. However, they did not consider the nonlocality or energy broadening of the SR scattering rates induced by the electric fields.
}%
Since the coupling strength of SR scattering is proportional to $\Delta \left( \mathbf{r}  \right) $, this approach is often referred to as the `linear' model and 
has successfully explained the effective electric field dependence of electron mobility under the strong inversion regimes.


%
\begin{table}[t]
\begin{center}
\caption{Roughness parameters from TEM measurements and theory}
\begin{tabular}{cccc}
\hline 
%
${ }$ & $ \Delta_{\rm A}$ (nm) & $\Lambda _{\rm A}$ (nm)  & Spectrum  \\
\hline
~Experiments (TEM) ~ & ~0.14--0.2~ & ~0.6--2.5~  & Gaussian~~~\cite{Goodnick1985} \\
~ ~ & ~0.2~ & ~1.5~  & Gaussian~~\cite{Zhao2010} \\
\hline
~Theory (linear model) ~ & ~0.44~ & ~2.5~  & Gaussian~~~\cite{Jungemann1993} \\
~ ~ & ~0.51~ & ~1.0~  & Gaussian~~\cite{Esseni2004} \\
~ ~ & ~0.55~ & ~1.3~  & Gaussian~~\cite{Ishihara2006} \\
%
\hline 
\end{tabular}
\label{Table_I}
\end{center}
\end{table}
Nevertheless, a long-time issue remains to exist: 
The roughness parameters employed in the linear model are usually very different from those measured by transmission electron microscopy (TEM), 
{\color{black}
as shown in Table~\ref{Table_I}. 
}%
Recently, this problem has been tackled by introducing nonlinear effects into the transition matrix elements of SR scattering, and realistic electron mobility has been successfully reproduced with the roughness parameters comparable to those from experiments~\cite{Lizzit2014,Badami2016,Lizzit2017}. These studies are motivated by the fact that spatial derivatives of the electrostatic potential and subband wave-function are discontinuous at the semiconductor/dielectric interface, which leads to ambiguity of the interface position. As pointed out in~\cite{Lizzit2014}, such ambiguity results from the difference in the dielectric constants and effective masses of the substrates on both sides of the interface.  

{\color{black}
In the present paper, 
}%
we investigate the above-mentioned discrepancy issue by taking an alternative approach. 
First, it should be noted that $\Delta \left( \mathbf{r}  \right)$ represents a `macroscopic' landscape of roughness which captures not only the short-range (scattering) components but also the long-range (drift) components over the entire interface plane~\cite{Sano2021PRE}. 
On the other hand, SR scattering is highly short-ranged, and thus, it would be more appropriate to treat $\Delta \left( \mathbf{r}  \right)$ microscopically. 
%
Second, discontinuity of spatial derivatives of the electrostatic potential at the interface simply implies that the Taylor expansion of the potential in powers of $\Delta \left( \mathbf{r}  \right)$ is invalid. 
Therefore, to sustain Ando's linear model in which the transition matrix due to SR scattering is proportional to $\Delta \left( \mathbf{r}  \right)$,  the stochastic nature of the interface position should be explicitly included. 
%
Third, SR scattering rates are usually calculated by employing Fermi's golden rule~\cite{Jacoboni2010book,Esseni2011}. 
However, the retarded self-energies due to short-range scattering are in many cases nonlocal (nondiagonal)~\cite{Doniach1998,Sano2025PRB}. Therefore, it is more preferable to use a fully quantum-mechanical framework to
derive the scattering rate.   

Based on these ideas, we construct a 
{\color{black}
stochastic and atomic-scale 
}%
SR model by introducing a probability density of roughness position at every atomic site in the interface plane. The roughness parameters in our model are turned out to be fully consistent with those from experiments. 
Furthermore, we derive the {\em self-consistent} SR scattering rate under the Green's function scheme, 
and it is shown 
that the self-consistency becomes important in determining realistic electron mobility 
under the regimes of strong effective electric field and/or low electron energy. 


\section{SR scattering models}
\label{sec:microtheory} 
Here, we briefly outline Ando's model of SR scattering in the inversion layer of Si-MOSFETs. A  
{\color{black}
stochastic atomic-scale 
}%
model proposed in the present paper 
and 
{\color{black}
the transition matrix elements expressed by the generalized Prange-Nee expression~\cite{Esseni2011,Prange1968}
}%
are then explained. 

\subsection{Conventional Ando's SR model}
\label{sec:convmodel}
%
The scattering potential in Ando's model is given by the spatial shift of the potential barrier $V_{{\rm ox}}$ at the interface between the semiconductor substrate and the dielectric~%
{\color{black}
\cite{Ando1977}. 
}%
The electrostatic Hartree potential in the substrate is then induced by the charge sheet at the interface due to ionized impurities in the substrate and by the one at the middle of the inversion layer due to electrons generated under strong inversion. The change of electric field in the substrate due to the presence of SR is usually ignored. 
Then, the scattering potential due to SR is simply given by
\begin{align}
{V_{{\rm sr}}}\left( {{\mathbf{r}},z} \right) = {V_{{\rm ox}}}\left[ {\theta \left( { - z + \Delta \left( {\mathbf{r}} \right)} \right) - \theta \left( { - z} \right)} \right] ,
\label{eq:Vsr}
\end{align}
where $z$ is the distance from the unperturbed interface, ${\mathbf{r}}$ is a 2-D position vector of roughness on the interface plane, and $\theta \left( {z} \right)$ is the step function. 
Expanding $V_{{\rm sr}}$ around $z=0$ and leaving only the linear term
{\color{black}
\cite{Ando1982}, 
}%
we obtain 
\begin{align}
{V_{{\rm sr}}}\left( {{\mathbf{r}},z} \right) \approx  {V_{{\rm ox}}}  \Delta \left( {\mathbf{r}} \right)  \delta \left( { z} \right) .
\label{eq:Vsr2}
\end{align}
{\color{black}
$V_{sr}$ is treated as the SR scattering potential and the dielectric screening could be implemented by dividing $V_{sr}$ by the dielectric function~\cite{Jin2007}. We would like to point out, however, 
}%
that the interface position of the unperturbed plane is strictly fixed at $z=0$ in Eq.~\eqref{eq:Vsr2}. 
{\color{black}
This is, in most cases, inappropriate due to the different dielectric constants of the substrates on both sides of the interface. Hence, 
}%
in the nonlinear model
{\color{black}
\cite{Lizzit2014}, 
}%
the exact expression of Eq.~\eqref{eq:Vsr} is employed without invoking the Taylor expansion. In the present study, nevertheless, we employ Eq.~\eqref{eq:Vsr2} because of its physical transparency and simplicity. 

The autocorrelation function of SR, 
{\color{black}
which is used for the calculations of the transition materix elements in \ref{sec:matrixelement},
}%
is defined by 
\begin{align}
C_{\rm A}\left( \mathbf{r}  \right) = \frac{1}{S}\int_S {{d^2}r'\Delta \left( {{\mathbf{r'}} + \frac{\mathbf{r} }{2}} \right)\Delta \left( {{\mathbf{r'}} - \frac{\mathbf{r} }{2}} \right)} ,
\label{eq:Gnnp}
\end{align}
where $S$ is the area of the interface. 
It should be noted that this definition is equivalent to taking the spatial average of the correlation between roughness and, thus, implicitly assuming 
translational invariance in space. This point becomes of critical importance in any quantum-mechanical treatment of short-range scattering under inhomogeneous device structures~\cite{Sano2025PRB,Sano2025TED}.
In the literature, $C_{\rm A}\left( \mathbf{r} \right)$ is modeled either by the Gaussian
\begin{align}
{C_{\rm A}}\left( \mathbf{r}  \right) = {\Delta _{{\rm A}}}^2\exp \left[ { - \frac{{{\mathbf{r} ^2}}}{{{\Lambda _{{\rm A}}}^2}}} \right]
\label{eq:Cand1}
\end{align}
or the exponential 
\begin{align}
{C_{\rm A}}\left( {\mathbf{r}} \right) = {\Delta _{{\rm A}}}^2\exp \left[ { - \frac{{\sqrt 2 \left| {\mathbf{r}} \right|}}{{{\Lambda _{{\rm A}}}}}} \right] ,
\label{eq:Cand2}
\end{align}
where $\Delta _{{\rm A}}$ is the deviation of the interface and $\Lambda _{{\rm A}}$ is the correlation length of roughness.
The roughness parameters ($\Delta _{{\rm A}}$ and $\Lambda _{{\rm A}}$) measured by TEM show a large discrepancy from the theoretical values found under the linear model, as summarized for the case of the Gaussian spectrum in Table~\ref{Table_I}. 

\subsection{Stochastic atomic-scale SR model}
\label{sec:micromodel}
We would like to stress again that $\Delta \left( {\mathbf{r}} \right)$ in Eq.~\eqref{eq:Vsr} represents a landscape of SR over the entire interface plane, namely, a global (macroscopic) structure of SR. On the other hand, the roughness parameters ($\Delta _{{\rm A}}$ and $\Lambda _{{\rm A}}$) found from TEM measurements are of the order of a few angstroms and, thus, atomistic. Therefore, $\Delta \left( {\mathbf{r}} \right)$ should be modeled from the 
{\color{black}
atomic-scale 
}%
viewpoint. 

\begin{figure}[tb]%
 \begin{minipage}{1\hsize}
 \begin{center}
  \includegraphics[width=7cm]{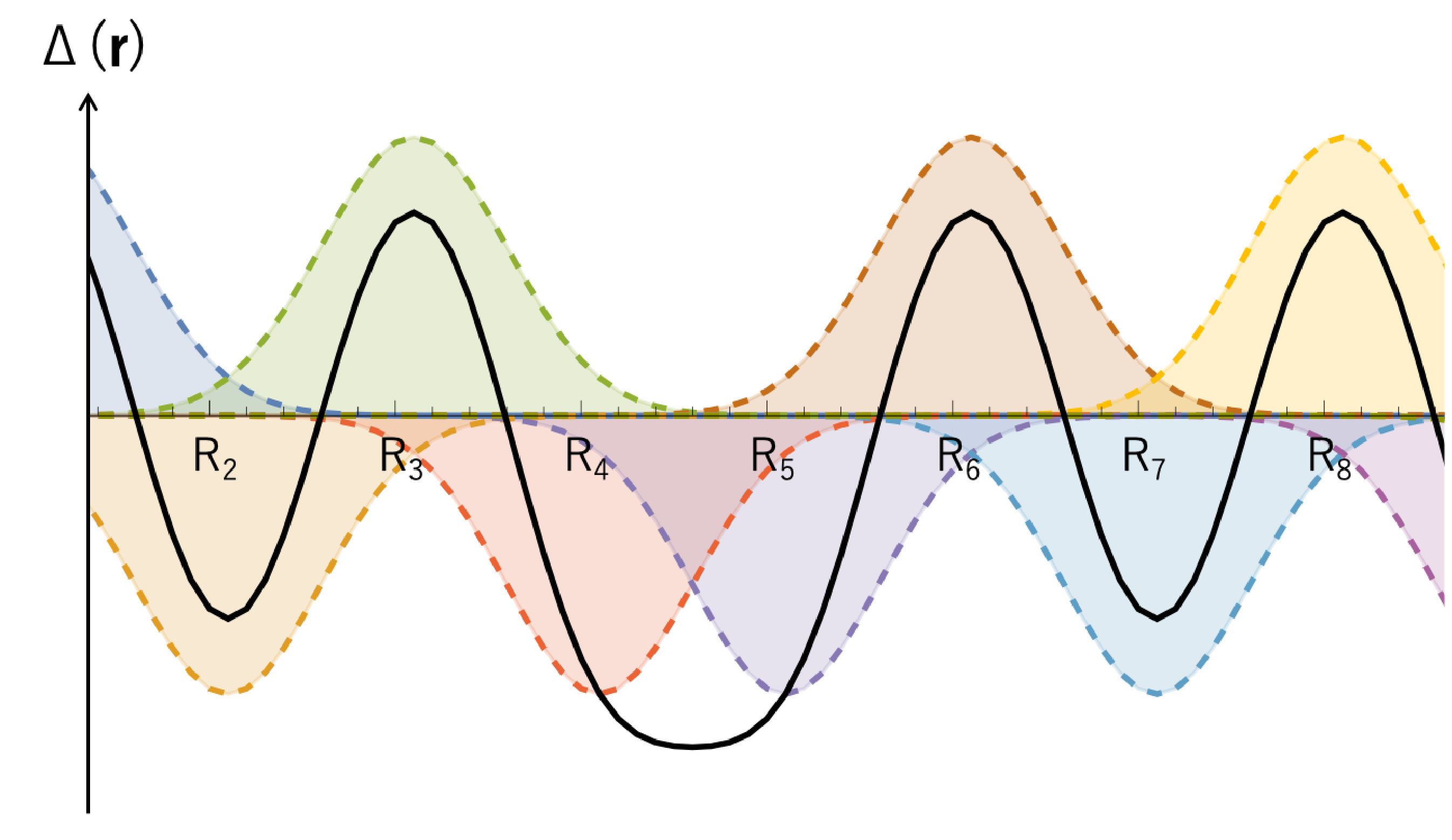}
  \caption{
 Schematic drawing of the roughness deviations $\Delta \left( {\mathbf{r}} \right)$ in the present model (dashed filled curves) and in Ando's model (thick black curve). 
The positions of atomic sites on the interface are represented by R$_{2}$ to R$_{8}$.
}
 \label{fig:Delta_r}
 \end{center}
 \end{minipage}
 \end{figure}

As mentioned in Sec.~\ref{sec:intro}, discontinuity of spatial derivatives of the potential at the semiconductor/dielectric interface invalidates the Taylor expansion of $V_{\rm sr}$. 
{\color{black}
In addition, the spatial derivative of the subband wave-functions is also discontinuous at the interface, which induces the magnitude of the SR transition matrix elements to be uncertain over the distance $\Delta \left( {\mathbf{r}} \right)$ (see Fig.~2 in \cite{Lizzit2014}). Hence, 
}%
a stochastic nature of the interface position should be explicitly included 
{\em as long as the linear model is sustained, where the interface position is fixed. 
} %
We introduce a probability density ${p_{X}}\left( {\mathbf{r}} \right)$ of finding an {\em atomistic} roughness at position $\mathbf{r}$ in the interface plane. Then, $\Delta \left( {\mathbf{r}} \right)$ is given by a simple sum of the roughness deviation {\em at every atomic site},  
and may be expressed by
\begin{align}
\Delta \left( {\mathbf{r}} ; \left\{ {{{\mathbf{R}}_l}} \right\} \right) = \sum\limits_{l = 1}^{{N_l}} {{{\bar \Delta} _{{\rm m}}}{{a}^2}{p_{X}}\left( {{\mathbf{r}} - {{\mathbf{R}}_l}} \right)} ,
\label{eq:mDelta}
\end{align}
where the magnitude of ${\bar \Delta} _{\rm m}$ 
is the standard deviation of $ \Delta \left( {\mathbf{r}} \right) $ 
{\color{black}
and its sign should be either positive or negative with equal probability, 
}%
$a$ is the mean separation of atoms in the interface plane, $N_l$ is the number of atoms on the interface, and ${\mathbf{R}}_l$ is the position of the $l$-th atom in the interface plane. 
{\color{black}
Since ${\bar \Delta} _{\rm m}$ is constant at every atomic site, the SR scattering potential induced by Eq.~\eqref{eq:mDelta} has the same potential profile at each atomic site. 
Fig.~\ref{fig:Delta_r} shows a schematic drawing of the roughness deviations $\Delta \left( {\mathbf{r}} \right)$ in the present model and in Ando's model. 
}%

The autocorrelation function of SR is found to be
\begin{align}
{C_X}\left( {\mathbf{r}} \right) = \frac{{{{{\bar \Delta} _{\rm m}}}^2{{a}^2}}}{S}\sum\limits_{\mathbf{q}} {{e^{i{\mathbf{q}} \cdot {\mathbf{r}}}}{{\left| {{{\tilde p}_{X}}\left( {\mathbf{q}} \right)} \right|}^2}} ,
\label{eq:CX}
\end{align}
where ${{{\tilde p}_{X}}\left( {\mathbf{q}} \right)}$ is the Fourier transform of ${p_{X}}\left( {\mathbf{r}} \right)$ and the relation $S \approx N_l a^2$ was used.  
%
{\color{black}
In the process of deriving Eq.~\eqref{eq:CX}, the correlation among the roughness at different atomic sites is ignored. As a result, the roughness-position dependence $\left\{ {{{\mathbf{R}}_l}} \right\}$ is lost from Eq.~\eqref{eq:CX}.
}%
Hence, this approximation is equivalent to taking the spatial average of the roughness position over the interface plane (namely, self-averaging), and  
{\color{black}
the present model implicitly assumes self-averaging over the roughness position in the interface plane.  
}%

Assuming that ${p_{X}}$ is given by the (2-D) Gaussian with the standard deviation $\sigma _{\rm m}$,
\begin{align}
{p_{X}}\left( {\mathbf{r}} \right) = \frac{1}{{2\pi {{\sigma_{\rm m}}^2}}}\exp \left[ { - \frac{{{{\mathbf{r}}^2}}}{{2{{\sigma_{\rm m}}^2}}}} \right] ,
\label{eq:patm}
\end{align}
Eq.~\eqref{eq:CX} becomes 
\begin{align}
{C_X}\left( {\mathbf{r}} \right) = \frac{{{{\bar \Delta} _{{\rm m}}}^2} a^2}{{4\pi {\sigma_{\rm m}}^2}}\exp \left[ { - \frac{{{{\mathbf{r}}^2}}}{{4{{\sigma_{\rm m}}^2}}}} \right] .
\label{eq:CXatm}
\end{align}
{\color{black}
Notice that Eq.~\eqref{eq:CXatm} is given by the Gaussian and consistent with the autocorrelation function assumed by Eq.~\eqref{eq:Cand1}. 
}%
Comparing Eq.~\eqref{eq:CXatm} with Eq.~\eqref{eq:Cand1}, we find that the roughness parameters in the present microscopic model (${\bar \Delta} _{{\rm m}}$ and $\sigma_{\rm m}$) are expressed in terms of $\Delta _{{\rm A}}$ and $\Lambda _{{\rm A}}$, which are found from experiments. They are given by 
\begin{align}
{ {\bar \Delta} _{\rm m} }= \frac{{\sqrt {\pi}} {\Delta _{\rm A}} {\Lambda _{\rm A}}}{a} 
\quad {\text{and}}\quad \sigma_{\rm m} = \frac{{{\Lambda _{\rm A}}}}{2} .
\label{eq:roughpara}
\end{align}
Given the values found from TEM measurements (Table~\ref{Table_I}), 
theoretical roughness parameters evaluated by Eq.~\eqref{eq:roughpara} are expected to be ${{\bar \Delta} _{{\rm m}}} = 0.39 \sim 2.3$ nm and $\sigma_{\rm m} = 0.3 \sim 1.25$ nm, where we employed $a=0.38$ nm for Si (100). 
Notice that $\sigma_{\rm m}$ is a bit larger than the atomic spacing $a$, as expected, whereas ${\bar \Delta} _{\rm m}$ 
is much larger than the experimental value of $\Delta _{{\rm A}}$ and may vary over a broad range.

{\color{black}
\subsection{SR transition matrix elements}
\label{sec:matrixelement}
Assuming that the wave-functions in the semiconductor substrate parallel to the interface plane are given by plane waves,  
the square of the transition matrix element due to SR scattering is given by
\begin{align}
{\left| {{M_{nn'}}\left( {\mathbf{q}} \right)} \right|^2} & = {\left| {{\Xi _{nn'}}} \right|^2}\frac{1}{S}\int_S {{d^2}r{e^{ - i{\mathbf{q}} \cdot {\mathbf{r}}}}{C_{\rm A/X}}\left( {\mathbf{r}} \right)} 
\nonumber \\
& = {\left| {{\Xi _{nn'}}} \right|^2}\frac{1}{S}{{\tilde C}_{\rm A/X}}\left( {\mathbf{q}} \right) ,
\label{eq:Mnnp}
\end{align}
where ${{\tilde C}_{\rm A/X}}\left( {\mathbf{q}} \right)$ is the Fourier transform of 
the autocorrelation functions expressed by either \eqref{eq:Cand1} or \eqref{eq:CXatm}. 
The coupling constant $\Xi _{nn'}$ (energy per unit length) is given by the generalized Prange-Nee expression, namely, 
\begin{align}
 {\Xi _{nn'}} & = \frac{{{\hbar ^2}}}{{2m_{\perp}}}\frac{{d\zeta _n^*(0)}}{{dz}}\frac{{d{\zeta _{n'}}(0)}}{{dz}} \nonumber \\
& = \int_0^\infty  {dz  \, \zeta _n^*(z)\left( {\frac{{\partial V_H}}{{\partial z}}} \right){\zeta _{n'}}(z)} 
\nonumber \\
& ~~~ + \frac{{{\varepsilon _n} - {\varepsilon _{n'}}}}{2}\int_0^\infty  {dz\left( {\zeta _n^*\frac{{d{\zeta _{n'}}}}{{dz}} - \frac{{d\zeta _n^*}}{{dz}}{\zeta _{n'}}} \right)} 
\label{eq:Xinnp}
\end{align}
for single-gate FETs~\cite{Esseni2011,Prange1968}. 
Here, $m_{\perp}$ is the effective mass (perpendicular to the interface plane) of the semiconductor substrate, $\zeta _n$ is the $n$-th subband wavefunction in the inversion layer, ${\varepsilon _n}$ is the eigen-energy of the $n$-th subband, and $V_H$ is the Hartree potential (energy) inside the inversion layer. 
In the conventional approaches, Fermi's golden rule, along with taking a sum over the final subband $n'$ in Eq.~\eqref{eq:Mnnp}, is employed to evaluate the SR scattering rate.  As we shall show in \ref{sec:srGF}, this treatment is equivalent to a local (diagonal) approximation of the self-energy associated with SR scattering. 
}%

\section{SR scattering rate under Green's function scheme}
\label{sec:srGF}
According to our previous studies\cite{Sano2021PRE,Sano2025PRB}, short-range scattering in semiconductors usually becomes nonlocal in space under a quantum-mechanical framework such as the nonequilibrium Green's functions (NEGF) method. Here, 
{\color{black}
the retarded self-energy due to SR scattering with the Born approximation is employed to derive the scattering rate. A similar methodology has been employed in \cite{Kannan2017}, but we would focus on the nonlocality of the SR scattering rates and energy broadening induced by the electric field in inversion layers of MOSFETs.
}%

\subsection{General theory}
\label{sec:Gtheory}
{\color{black}
The scattering potential (operator) $V_{{\rm sr}}$ is given by Eq.~\eqref{eq:Vsr2}, in which $\Delta {\left( {\mathbf{r}} \right)}$ is now replaced by Eq.~\eqref{eq:mDelta}. 
}%
The retarded self-energy operator under the Born approximation is then expressed, in the 1st-quantized form, by 
\begin{align}
{{\hat \Sigma }^r}_{{\rm sr}} \left( E \right) = {V_{{\rm sr}}}{\left( {{\mathbf{\hat r}},\hat z;\left\{ {{{\mathbf{R}}_l}} \right\}} \right) }{{\hat G}^r}\left( E \right){V_{{\rm sr}}}\left( {{\mathbf{\hat r}},\hat z;\left\{ {{{\mathbf{R}}_l}} \right\}} \right) ,
\label{eq:selfsr}
\end{align}
where ${{\hat G}^r}$ is the retarded Green's operator and given by
\begin{align}
{{\hat G}^r}\left( E \right) = {\left[ {E - {{\hat H}_0} - {{\hat \Sigma }^r}_{{\rm sr}}\left( E \right)} \right]^{ - 1}}.
\label{eq:Gr}
\end{align}
The unperturbed Hamiltonian ${\hat H}_0$ is given by 
\begin{align}
{{\hat H}_0} = \frac{{{{{\mathbf{\hat p}}}^2}}}{{2m_{\parallel}}} + \left[ \frac{{{{\hat p}_z}^2}}{{2m_{\perp}}} + {V_H}\left( {\hat z} \right) \right] \equiv {{\hat h}_{\mathbf{p}}} + {{\hat h}_z} ,
\label{eq:H0}
\end{align}
where $\mathbf{\hat p}$ is the momentum operator in the interface plane, $\hat p_z$ is that in the $z$-direction (perpendicular to the interface), and
{\color{black}
$m_{\parallel}$ is the effective mass parallel to the interface plane. 
}%
The retarded self-energy in real-space representation is then derived as
\begin{align}
& \left\langle {{\mathbf{r}},z} \right|{{\hat \Sigma }^r_{{\rm sr}}}\left( E \right)\left| {{\mathbf{r'}},z'} \right\rangle  
\approx {V_{{\rm ox}}}^2\delta \left( z \right)\delta \left( {z'} \right){{\bar \Delta} _{\rm m}}^2{{a}^4}
\nonumber \\
& ~~ \times \sum\limits_{l = 1}^{{N_l}} {{p_{X}}\left( {{\mathbf{r}} - {{\mathbf{R}}_l}} \right){p_{X}}\left( {{\mathbf{r'}} - {{\mathbf{R}}_l}} \right)} \left\langle {{\mathbf{r}},z} \right|{{\hat G}^r}\left( E \right)\left| {{\mathbf{r'}},z'} \right\rangle  ,
\label{eq:selfsr2}
\end{align}
where the correlation among roughness at different atomic sites is ignored. This is also justified because $p_{X}$ is significant only in the region of each atomic site. 
Using the Wigner coordinates  in the interface plane defined by ${\mathbf{X}} = \left( {{\mathbf{r}} + {\mathbf{r'}}} \right)/2$ and ${\boldsymbol\xi}  = {\mathbf{r}} - {\mathbf{r'}}$, the left-hand-side of Eq.~\eqref{eq:selfsr2} is expressed by 
\begin{align}
\left\langle {{\mathbf{r}},z} \right|{{\hat \Sigma }^r}_{{\rm sr}}\left( E \right)\left| {{\mathbf{r'}},z'} \right\rangle  = \sum\limits_{s,s'} {\zeta _s^{}\left( z \right)\zeta _{s'}^*\left( {z'} \right)\Sigma _{{\rm sr},ss'}^r\left( {{\mathbf{X}},\boldsymbol\xi ;E} \right)}  ,
\label{eq:selflhs}
\end{align}
where the matrix element of ${{\hat \Sigma }^r}_{{\rm sr}} \left( E \right) $ is given by
\begin{align}
\Sigma _{{\rm sr},ss'}^r\left( {{\mathbf{X}},\boldsymbol\xi ;E} \right) \equiv \left\langle {{\mathbf{X}} + \frac{\boldsymbol\xi }{2},s} \right|{{\hat \Sigma }^r}_{{\rm sr}}\left( E \right)\left| {{\mathbf{X}} - \frac{\boldsymbol\xi }{2},s'} \right\rangle .
\end{align}
Consequently, we can derive the equation for finding the retarded self-energy due to SR scattering:
\begin{align}
& \Sigma _{{\rm sr},ss'}^r\left( {{\mathbf{X}},\boldsymbol\xi ;E} \right) = \sum\limits_{n,n'} {{\Xi _{sn}}} \sum\limits_{l = 1}^{{N_l}} {{{a}^2}\delta \left( {{\mathbf{X}} - {{\mathbf{R}}_l}} \right)}  
\nonumber \\
&~ \times \frac{{{{\bar \Delta} _{\rm m}}^2{{a}^2}}}{S}\sum\limits_{\mathbf{q}} {{e^{i{\mathbf{q}} \cdot \boldsymbol\xi }}{{\left| {{{\tilde p}_{X}}\left( {\mathbf{q}} \right)} \right|}^2}} G_{nn'}^r\left( {{\mathbf{X}},\boldsymbol\xi ;E} \right){\Xi _{n's'}} .
\label{eq:selfsc}
\end{align}
Here, we used Ando's asymptotic expression~\cite{Ando1982} of the subband wave-function near $z=0^{+}$; 
\begin{align}
\zeta _s^{}\left( z \right) \simeq \frac{\hbar }{{\sqrt {2m_{\perp}{V_{{\rm ox}}}} }}\frac{{d\zeta _s^{}\left( 0 \right)}}{{dz}}\exp \left[ { - \frac{{\sqrt {2m_{\perp}{V_{{\rm ox}}}} }}{\hbar }z} \right] .
\label{eq:Andowf}
\end{align}
{\color{black}
Notice that $\Sigma _{{\rm sr},ss'}^r\left( {{\mathbf{X}},\boldsymbol\xi ;E} \right)$ is intrinsically nonlocal with respect to positions and subband indices~\cite{Sano2021PRE,Sano2025PRB}. 
}%
Assuming that the roughness position in the interface plane is distributed uniformly and randomly, we take the spatial average of the roughness position over the plane. The self-averaged retarded self-energy is then given by
\begin{align}
\left\langle {\Sigma _{{\rm sr},ss'}^r\left( {{\mathbf{X}},\boldsymbol\xi ;E} \right)} \right\rangle  = \sum\limits_{n,n'} {{\Xi _{sn}} {{C_X}\left( \boldsymbol\xi  \right)G_{nn'}^r\left( {{\mathbf{X}},\boldsymbol\xi ;E} \right)} {\Xi _{n's'}}}  ,
\label{eq:selfsrav}
\end{align}
where Eq.~\eqref{eq:CX} was used. 
Notice that $G_{nn'}^r$ is approximated by Eq.~\eqref{eq:Gr}, in which ${{{\hat \Sigma }^r}_{{\rm sr}}}$ is replaced by 
${\left\langle {\hat \Sigma _{{\rm sr}}^r} \right\rangle }$. 
Hence, Eq.~\eqref{eq:selfsrav} needs to be solved in a self-consistent manner. Fourier-transforming both sides 
with respect to $\boldsymbol\xi$, an alternative expression of the retarded self-energy is obtained as 
\begin{align}
& \left\langle {\Sigma _{{\rm sr},ss'}^r\left( {{\mathbf{X}},{\mathbf{k}};E} \right)} \right\rangle 
\nonumber \\
&~ = \sum\limits_{n,n'} {{\Xi _{sn}}} \left[ {\frac{1}{S}\sum\limits_{\mathbf{q}} {{{\tilde C}_X}\left( {\mathbf{q}} \right)G_{nn'}^r\left( {{\mathbf{X}},{\mathbf{k}} - {\mathbf{q}};E} \right)} } \right]{\Xi _{n's'}} ,
\label{eq:selfsravk}
\end{align}
where ${\tilde C}_X \left( {\mathbf{q}} \right)$ is expressed by
\begin{align}
{{\tilde C}_X}\left( {\mathbf{q}} \right) = {{\bar \Delta} _{\rm m}}^2{{a}^2}{\left| {{{\tilde p}_{X}}\left( {\mathbf{q}} \right)} \right|^2} .
\label{eq:CXq}
\end{align}
The SR scattering rate for electrons is now obtained by taking the imaginary part of Eq.~\eqref{eq:selfsravk};  
\begin{align}
\frac{1}{{{\tau _{{\rm sr},ss'}}\left( {{\mathbf{X}},{\mathbf{k}};E} \right)}} =  - \frac{2}{\hbar }\operatorname{Im} \left[ {\left\langle {\Sigma _{{\rm sr},ss'}^r\left( {{\mathbf{X}},{\mathbf{k}};E} \right)} \right\rangle } \right] .
\label{eq:sr_rate}
\end{align}
Eqs.~\eqref{eq:selfsrav} and \eqref{eq:selfsravk} are one of the main results of the present paper: They are the formulas to determine the self-energy due to SR scattering in a self-consistent manner. 

We would like to stress again that the scattering rate given by Eq.~\eqref{eq:sr_rate} is nondiagonal with respect to $s$ and $s'$. This is reasonable because the interference between two different quantum states must be involved in any calculations of transport properties. 
Unfortunately, the past studies of SR scattering have completely missed this point. 
The reason why Eq.~\eqref{eq:sr_rate} is diagonal in $\mathbf{k}$ is that it carries the center-of-mass coordinate $\mathbf{X}$. 
In the present study, however, the self-averaging and bulk approximation have been employed to derive Eqs.~\eqref{eq:selfsrav} and \eqref{eq:selfsravk} so that they are essentially independent of $\mathbf{X}$. Nevertheless, when the present scheme is applied to inhomogeneous device structures, 
the retarded self-energy becomes dependent on $\mathbf{X}$ 
even after self-averaging. This fact has already been demonstrated in the cases of impurity scattering, in which the discrete nature of impurities doped in the substrate is taken into account so that the impurity scattering becomes nonlocal in space even under uniform impurity configurations~\cite{Sano2025PRB,Sano2025TED}.

{\color{black}
Also, it should be noted that the screening effect is intentionally ignored in the present model. Since the length-scale of the scattering potential is much smaller than the screening length, screening effects are insignificant in the present model. Indeed, screening would be crucial in the conventional Ando's model because the scattering potential induced by a macroscopic $\Delta \left( \mathbf{r} \right) $ contains the long-range components which must be eliminated by screening.
}%

\subsection{Reduction to the formula from Fermi's golden rule}
\label{sec:FermiGR}
To obtain the conventional formula derived from Fermi's golden rule, $G_{nn'}^r$ in Eq.~\eqref{eq:selfsravk} is replaced by the unperturbed retarded Green's function 
$g_{nn'}^r$, which is defined by
\begin{align}
g_{nn'}^r\left( {{\mathbf{k}};E} \right) = \frac{1}{{E + i{0^ + } - {\varepsilon _{\mathbf{k}}} - {\varepsilon _n}}}{\delta _{n,n'}} ,
\label{eq:grnn}
\end{align}
where $\varepsilon _{\mathbf{k}}$ is the electron energy with wave-vector $\mathbf{k}$ associated with $\hat h_{\mathbf{p}}$. 
Considering the diagonal component, 
Eq.~\eqref{eq:sr_rate} becomes
\begin{align}
\frac{1}{{{\tau _{{\rm sr},ss}}\left( {{\varepsilon _{\mathbf{k}}}} \right)}} & = \frac{{2\pi }}{\hbar }\sum\limits_n {{{\left| {{\Xi _{ns}}} \right|}^2}} 
\nonumber \\
& \times \frac{1}{S}\sum\limits_{\mathbf{q}} {{{\tilde C}_X}\left( {\mathbf{q}} \right)\delta \left( {{\varepsilon _{\mathbf{k}}} - {\varepsilon _{{\mathbf{k}} - {\mathbf{q}}}} - \delta {\varepsilon _{ns}}} \right)} 
\label{eq:Fgrule}
\end{align}
where we set $E = {\varepsilon _{\mathbf{k}}} + {\varepsilon _s}$ and $\delta {\varepsilon _{ns}} = {\varepsilon _n} - {\varepsilon _s}$. 
%
Assuming a parabolic band structure for electrons and the Gaussian probability density ${p_{X}}\left( \mathbf{r} \right)$ expressed by Eq.~\eqref{eq:patm}, Eq.~\eqref{eq:Fgrule} is found to be 
\begin{align}
\frac{1}{{{\tau _{{\rm sr},ss}}\left( {z} \right)}} & = \frac{m_{\parallel}}{{{\hbar ^3}}}{{\bar \Delta} _{\rm m}}^2{{a}^2}\sum\limits_n {{{\left| {{\Xi _{ns}}} \right|}^2}} 
\nonumber \\
&~~ \times {e^{ - \left( {z - \frac{{\delta {z_{ns}}}}{2}} \right)}}{I_0}\left[ {\sqrt {z\left( {z - \delta {z_{ns}}} \right)} } \right] ,
\label{eq:Fgrule_para}
\end{align}
where $I_0$ is the modified Bessel function of the $0$-th order, $z = {\varepsilon _{\mathbf{k}}}/{\varepsilon _\Lambda }$ with ${\varepsilon _\Lambda } = {\hbar ^2}/\left( {4m_{\parallel}{\sigma_{\rm m}^2}} \right)$, and 
$\delta {z_{ns}} = \delta {\varepsilon _{ns}}/{\varepsilon _\Lambda }$. 
Given the relationship of the roughness parameters expressed by Eq.~\eqref{eq:roughpara}, we can immediately recognize that Eq.~\eqref{eq:Fgrule_para} is identical to the formula derived from Fermi's golden rule in the linear model~\cite{Goodnick1985}. 
%
%

\section{Numerical Results and Discussion}
\label{sec:R&D}
%
The present model is applied to the inversion layer in bulk Si-MOSFETs. Application of the present scheme to more complicated device structures such as UTB-MOSFETs is, in principle, straightforward. 
Since our interests lie in the impact of SR scattering on transport properties, most attention is focused on the strong inversion regimes 
and, thus, only the lowest subband ($n=1$) is included in the following calculations. 
{\color{black}
Therefore, all numerical calculations here are based on the local (diagonal) approximation. More systematic analysis on nonlocality of the self-energy due to SR scattering requires full NEGF simulations~\cite{Mahan2000,Datta2005,Haug2008} and this part is left for future study.
}%

\subsection{Self-consistent scattering rates}
\label{sec:scscatrate}
\begin{figure}[tb]%
 \begin{minipage}{1\hsize}
 \begin{center}
  \includegraphics[width=9cm]{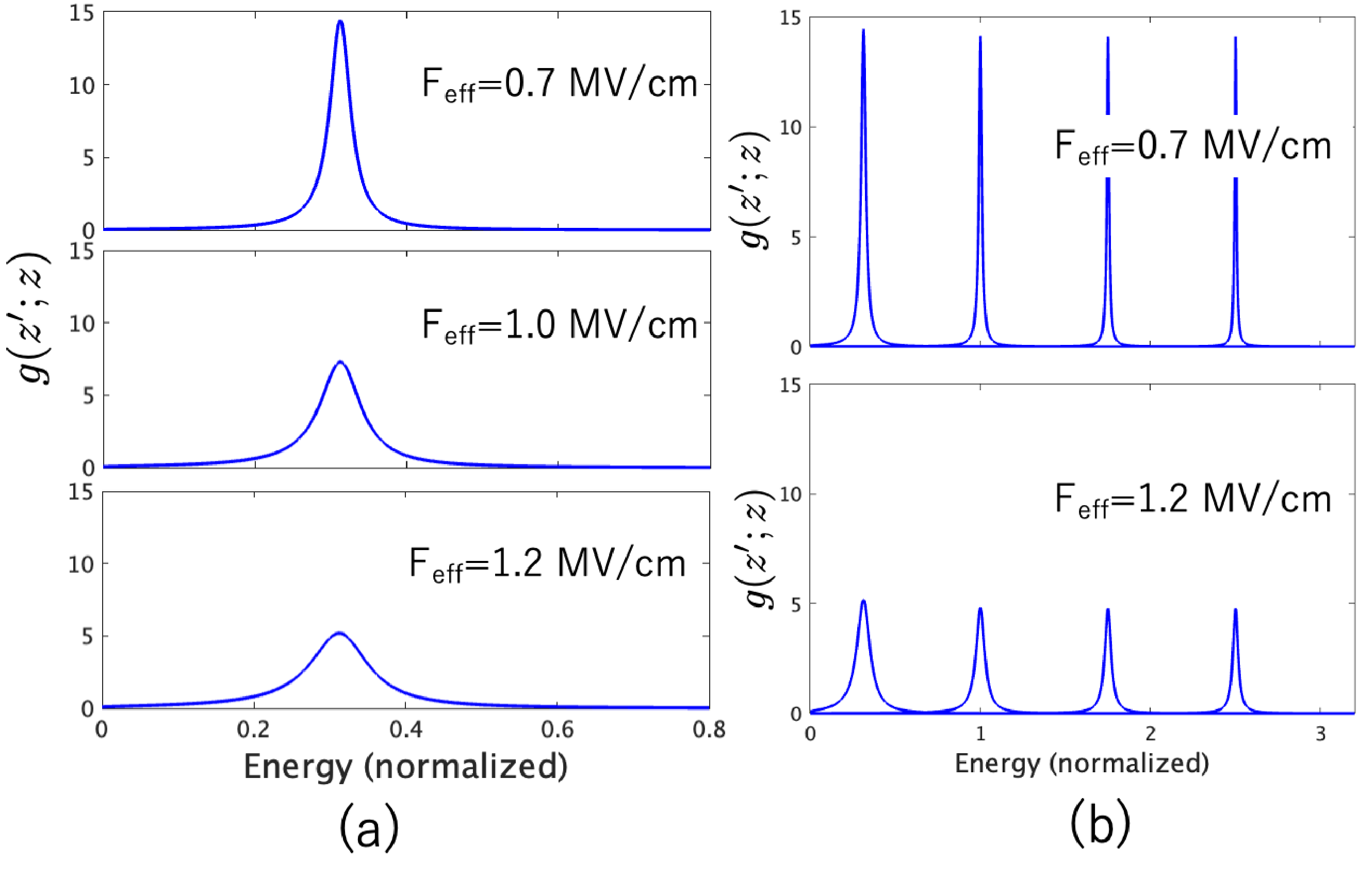}
  \caption{
  (a) Integrand $g\left( z' ; z \right) $ of Eq.~\eqref{eq:sr_rate00} at $z=$0.31 as a function of normalized electron energy $z'$ for three different effective electric fields, $F_{\rm eff} =$ 0.7, 1.0, 1.2 MV/cm. Electron energy is normalized as $z' = {\varepsilon _{\mathbf{k'}}}/{\varepsilon _\Lambda }$. 
  The roughness parameters  of ${\bar \Delta}_{\rm m} = $ 1.3 nm and $\sigma_{\rm m} =$ 0.7 nm are assumed. 
  (b) $g\left( z' ; z \right) $ as a function of normalized electron energy $z'$ at $z =$ 0.31, 1, 1.75, 2.5 for $F_{\rm eff} =$ 0.7 and 1.2 MV/cm.
} 
  \label{fig:intgrd}
 \end{center}
 \end{minipage}
 \end{figure}%
Under the strong inversion approximation, the effective electric field $F_{\rm eff}$ in the inversion layer directly relates with the SR coupling constant, as seen from the second line of Eq.~\eqref{eq:Xinnp}; namely, ${eF_{\rm eff}} = \left| {{\Xi _{11}}} \right|$, where $e$ is the magnitude of electron charge. 
The SR scattering rate is then calculated by Eqs.~\eqref{eq:selfsravk} and \eqref{eq:sr_rate}, and it is given by
\begin{align}
{\gamma _1}\left( z \right) & = \frac{m_{\parallel}}{{{\hbar ^2}}}\frac{{{{\bar \Delta} _{\rm m}}^2{a}^2{(e F_{\rm eff})}^2}}{{2{\varepsilon _\Lambda }}}\int_0^\infty  {dz'{e^{ - \frac{{z + z'}}{2}}}{I_0}\left( {\sqrt {zz'} } \right)}  
\nonumber \\
&~~~~\times \frac{1}{\pi }\frac{{{\gamma _1}\left( {z'} \right)}}{{{{\left( {z - z'} \right)}^2} + {{\left[ {{\gamma _1}\left( {z'} \right)} \right]}^2}}} ,
\label{eq:sr_rate00}
\end{align}
where $z' = {\varepsilon _{\mathbf{k'}}}/{\varepsilon _\Lambda }$ and ${\gamma _1}$ is defined by
\begin{align}
{\gamma _1}\left( z \right) = 
- \frac{{\operatorname{Im} \left[ {\left\langle {\Sigma _{{\rm sr},11}^r\left( {E = {\varepsilon _{\mathbf{k}}} + {\varepsilon _1}} \right)} \right\rangle } \right]}}{{{\varepsilon _\Lambda }}}
=\frac{\hbar}{2 {\varepsilon _\Lambda }} 
\frac{1}{{{\tau _{{\rm sr},11}}\left( {\varepsilon _{\mathbf{k}}} \right)}}. 
\label{eq:gam0}
\end{align}
Notice that Eq.~\eqref{eq:sr_rate00} properly reduces to Eq.~\eqref{eq:Fgrule_para} as $\gamma _{1}$ becomes much smaller than the (normalized) energy separation $\left| z-z' \right|$.

The integrand of Eq.~\eqref{eq:sr_rate00} with the self-consistent ${\gamma _1}$, denoted by $g(z';z)$, is plotted in Fig.~\ref{fig:intgrd} for various effective electric fields and electron energies. 
{\color{black}
Here, $m_{\parallel} =0.19 m_0$ with $m_0$ being the electron's bare mass was used. 
}%
The theoretical roughness parameters of ${\bar \Delta} _{\rm m} =  1.3$ nm and $\sigma_{\rm m} =0.7$ nm, as well as $a=0.38$ nm, were used. 
In the conventional model where Fermi's golden rule is employed, the integrand is simply proportional to the delta-function, $\delta \left( z-z' \right)$, whereas in the present model, the delta-function is broadened and the magnitude of broadening gets larger as the effective field becomes larger, as shown in Fig.~\ref{fig:intgrd}~(a). On the other hand, the broadening gets smaller as the electron energy increases even under large effective fields, as shown in Fig.~\ref{fig:intgrd}~(b). Such characteristics reflect the fact that the magnitude of the scattering rate is large under the regimes of large effective fields and/or small electron energies so that time-energy uncertainty becomes most significant.

\begin{figure}[tb]%
 \begin{minipage}{1\hsize}
 \begin{center}
  \includegraphics[width=9cm]{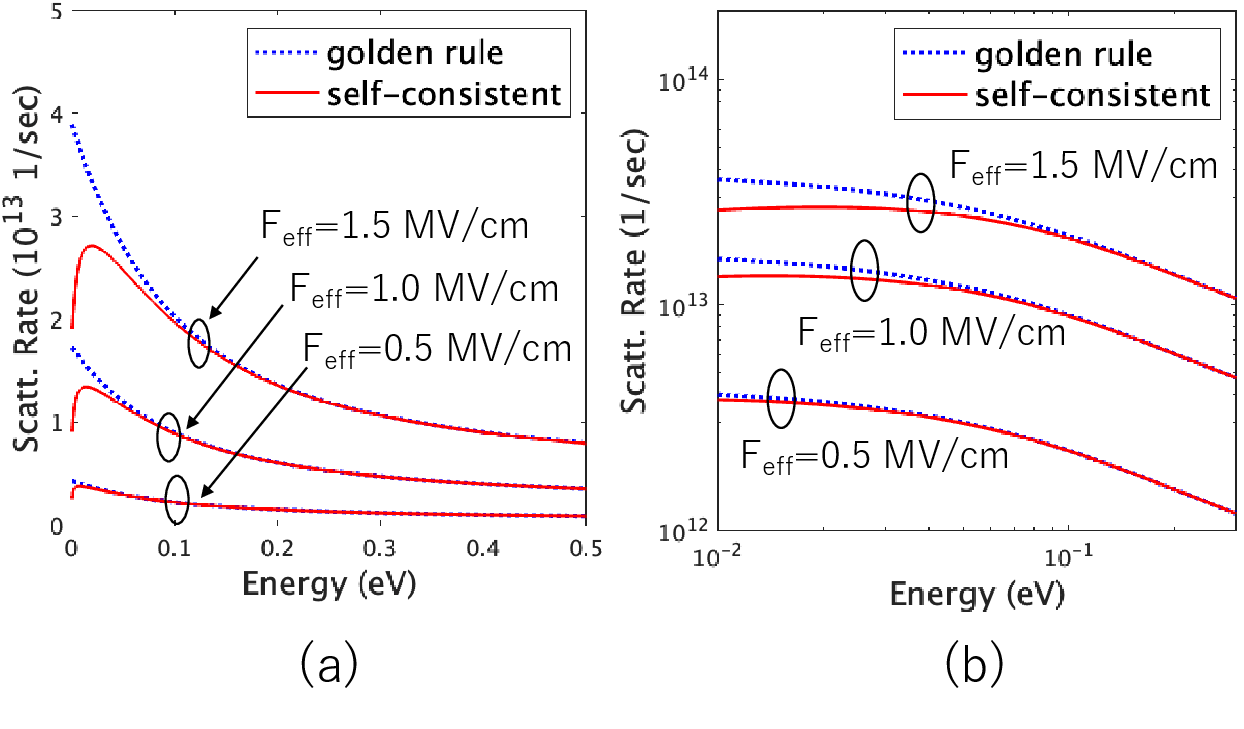}
  \caption{
 (a) SR scattering rates in Si obtained by the self-consistent scheme (red curves) and by Fermi's golden rule (blue dotted curves) as a function of electron energy for $F_{\rm eff} = $ 0.5, 1.0, and 1.5 MV/cm. The roughness parameters of ${\bar \Delta}_{\rm m} = $ 1.3 nm and $\sigma_{\rm m} =$ 0.7 nm are assumed.
 (b) Same as (a) on the log-log scale.
}
 \label{fig:SRrate}
 \end{center}
 \end{minipage}
 \end{figure}%
Figure~\ref{fig:SRrate} shows the SR scattering rates in Si obtained from the self-consistent scheme (red curves) and Fermi's golden rule (blue dotted curves) as a function of electron energy under $F_{\rm eff} = $ 0.5, 1.0 and 1.5 MV/cm. The self-consistent results greatly deviate from the conventional results (Fermi's golden rule) at low electron energies. In addition, this difference is more significant under large effective fields. These features are fully consistent with the observations of Fig.~\ref{fig:intgrd}. 
{\color{black}
Since transport characteristics under strong effective electric fields in inversion layers of MOSFETs are dominated by SR scattering with low-energy electrons, energy broadening obtained from the self-consistent calculations would be of critical importance for accurate predictions of mobility. 
}%

\subsection{SR-limited electron mobility in Si}
\label{sec:srmobility}
The transport properties are calculated under the quasi-equilibrium condition so that the electron distribution function is assumed to be the Fermi-Dirac distribution.
Then, the SR-limited electron mobility in Si is approximately evaluated by 
\begin{align}
{\mu _{\rm  sr}} 
= \frac{e}{m_{\parallel}}{\left[ {\frac{1}{{{N_s}}}\frac{m_{\parallel}}{{\pi {\hbar ^2}}}\int_0^\infty  {dE\frac{1}{{{e^{\frac{{E - {\mu _c}}}{{{k_B}T}}}} + 1}}\frac{1}{{{\tau _{sr,11}}\left( E \right)}}} } \right]^{ - 1}}
\label{eq:srmob}
\end{align}
where $\mu _c$ is the chemical potential, and $N_{\rm s}$ is the electron sheet density in the inversion layer. $N_{\rm s}$ is found from the usual relationship between the effective electric field and the sheet densities in the inversion layer, and thus, it is expressed by 
\begin{align}
{F_{\rm eff}} = \frac{e}{\epsilon _{\rm s}}\left( {{N_{\rm dep}} + \frac{{{N_{\rm s}}}}{2}} \right),
\label{eq:Feff}
\end{align}
where $\epsilon _{\rm s}$ is the dielectric constant of Si, and  $N_{\rm dep}$ is the dopant sheet density and assumed to be $N_{\rm dep} = 3 \times 10^{11}$ cm$^{-2}$ throughout this study.

\begin{figure}[tb]%
 \begin{minipage}{1\hsize}
 \begin{center}
  \includegraphics[width=9cm]{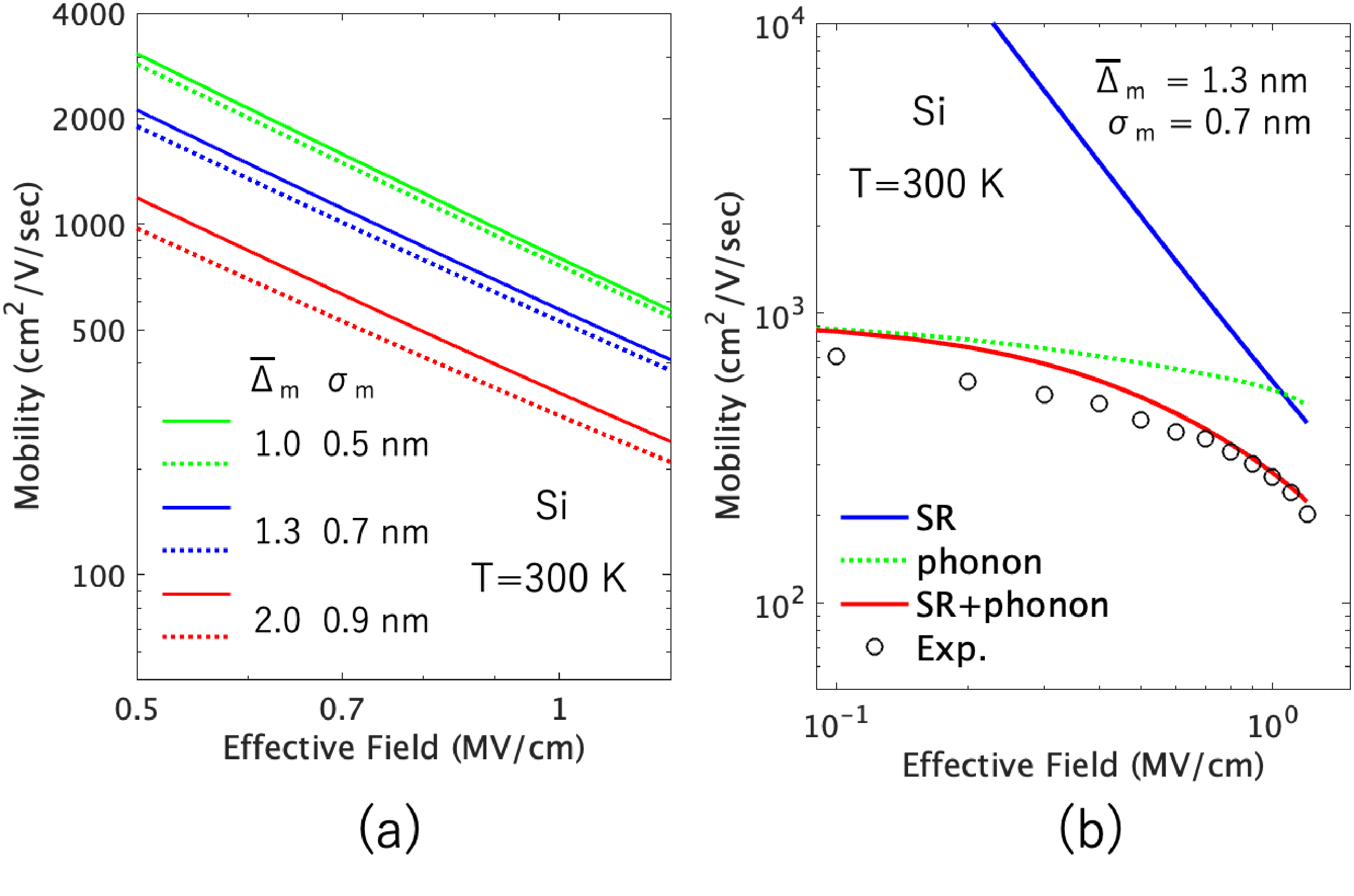}
  \caption{%
 (a) SR-limited electron mobility in Si as a function of effective electric field for various values of theoretical roughness parameters (Green, Blue, Red lines). Solid and dotted lines represent the mobilities obtained from the self-consistent SR scattering rates and from Fermi's golden rule, respectively. 
  (b) Electron mobilities in Si due to SR scattering (solid blue curve), phonon scattering (dotted green curve), and total (SR and phonon) scattering (solid red curve) as a function of effective field.  The roughness parameters  of ${\bar \Delta}_{\rm m} = $ 1.3 nm and $\sigma_{\rm m} =$ 0.7 nm are assumed. The phonon-limited mobility was taken from \cite{Jin2007}, and the experimental mobilities found from TEM are represented by open circles~\cite{Takagi1994}.
  } 
  \label{fig:mob}
 \end{center}
 \end{minipage}
 \end{figure}%
%
The roughness-parameter dependence of the SR-limited mobilities in Si, obtained from the self-consistent SR scattering rates and from the scattering rate through Fermi's golden rule, is shown as a function of effective electric field in Fig.~\ref{fig:mob}~(a). 
Compared with the results from Fermi's golden rule, the self-consistent scattering rate always yields larger mobility due to the reduction of the scattering rates at low electron energies. 
This difference gets larger as the roughness parameters (equivalently, the SR scattering rates) become larger.
{\color{black}
Also, we should mention that the calculation results are quite sensitivity to the values of the parameters, as seen in Fig.~\ref{fig:mob}~(a). 
}%
Electron mobilities due to SR scattering, phonon scattering, and total (SR and phonon) scattering are shown in Fig.~\ref{fig:mob}~(b). 
{\color{black}
The extraction of the roughness parameters is done by fitting the total mobility with that from experiments~\cite{Takagi1994}.  Since there are only two parameters in the present model, the parameters are uniquely determined.
}%
We find that the roughness parameters of ${\bar \Delta} _{\rm m} = 1.3$ nm and $\sigma_{\rm m} = 0.7$ nm reproduce a reasonable fit to the experimental mobility in~\cite{Takagi1994}. These values correspond to the experimental roughness parameters of $\Delta_{\rm A} = 0.2$ nm and $\Lambda_{\rm A} = 1.4$ nm, 
and they are fully consistent with the experimental results. 

Finally, we should mention that the roughness parameter $\sigma _{\rm m}$ in the present model reflects an uncertainty of roughness position, induced by the difference in dielectric constants of the materials on both sides of the interface. Hence, the difference in dielectric constants might be an important clue to understand the material dependence of the SR-limited mobility. 
In addition, the retarded self-energy due to SR scattering is nondiagonal with respect to subband indices, $s$ and $s'$. 
Therefore, more accurate calculations of transport properties should be carried out by a fully quantum-mechanical scheme such as the NEGF method, in which the energy dispersion of electrons determined by the retarded Green's function is self-consistently coupled with the nonequilibrium electron distribution function obtained from the lesser Green's function~\cite{Sano2025PRB,Datta2005,Haug2008}. 
{\color{black}
For such simulations, matrix multiplication with nondiagonal scattering matrices is mandatory, and more computational cost is demanded. However, it would not be critical because the nonzero elements of the scattering matrix dampen as the elements are distant from the diagonal. 
}%
The research along this direction is in progress and will be reported elsewhere.

\section{Conclusion}
\label{sec:concl}
We have proposed a 
{\color{black}
stochastic atomic-scale model 
}%
of SR scattering in inversion layers of bulk-MOSFETs. The stochastic nature of roughness position, which is induced by the discontinuity of spatial derivatives of the electrostatic potential at the interface, has been introduced at each atomic site with the probability density of roughness position. 
The SR scattering rate has been derived by the Green's function scheme, and we have shown that it is intrinsically nonlocal (nondiagonal) with respect to subband indices and positions. Furthermore, the self-consistent scattering rates deviate from the conventional results based on Fermi's golden rule and the deviation becomes significant under the strong inversion regimes. 
{\color{black}
The electron mobility has been estimated by the local (diagonal) approximation under the quasi-equlibrium conditions. 
}%
We have found that the roughness parameters in the present model are consistent with those from experiments, and thus, there is no discrepancy between theory and experiment as long as SR is treated `microscopically.'

%

%

\begin{thebibliography}{1}

\bibitem{Sun1980}
S.~C.~Sun and J.~D.~Plummer, ``Electron mobility in inversion and accumulation layers on thermally oxidized silicon surfaces,'' \emph{IEEE Trans. Electron Devices}, vol.~27, no.~8, pp. 1497-1508, Aug. 1980, 
doi: 10.1109/T-ED.1980.20063

\bibitem{Goodnick1985}
 S.~M.~Goodnick, D.~K.~Ferry, C.~W.~Wilmsen,Z.~Liliental, D~Fathy, and O.~L~Krivanek, ``Surface Roughness at the Si(100)-${\mathrm{SiO}}_{2}$ interface,'' 
\emph{Phys. Rev. B}, vol.~32, no.~12, 8171--8186, Dec. 1985, 
doi: 10.1103/PhysRevB.32.8171.

\bibitem{Takagi1988}
S.~Takagi, M.~Iwase and A.~Toriumi, ``On the universality of inversion-layer mobility in n- and p-channel MOSFETs,''  \emph{IEDM Tech. Dig.}, Dec. 1988, pp. 398--401, 
doi: 10.1109/IEDM.1988.32840.

\bibitem{Takagi1994}
S. Takagi, A. Toriumi, M. Iwase and H. Tango, ``On the universality of inversion layer mobility in Si MOSFET's: Part I-effects of substrate impurity concentration,'' \emph{IEEE Trans. Electron Devices}, vol.~41, no.~12, pp. 2357-2362, Dec. 1994, 
doi: 10.1109/16.337449.

\bibitem{Zhao2010}
Y.~Zhao, H.~Matsumoto, T.~Sato, S.~Koyama, M.~Takenaka, and S.~Takagi, ``A novel characterization scheme of Si/SiO2 interface roughness for surface roughness scattering-limited mobilities of electrons and holes in unstrained- and strained-Si MOSFETs,'' 
\emph{IEEE Trans. Electron Devices}, vol.~57, no.~9, pp.~2057--2066, Sep. 2010,
doi: 10.1109/TED.2010.2052394.

\bibitem{Ando1977}
T.~Ando, ``Screening effect and quantum transport in a silicon inversion layer in strong magnetic field,'' \emph{J. Phys. Soc. Jpn.}, vol.~43, no.~5, pp. 1616--1626,
  1977, 
  doi: 10.1143/JPSJ.43.1616.

\bibitem{Ando1982}
T.~Ando, A.~Fowler, and F. ~Stern, ``Electronic properties of two-dimen-
sional systems,'' \emph{Rev. Mod. Phys.}, vol.~54, no.~5, pp. 437--672,
  April 1982, 
  doi: 10.1103/RevModPhys.54.437.

\bibitem{Jin2007TED} 
S.~Jin, M.~V.~Fischetti, and T-w.~Tang,
``Modeling of Surface-Roughness Scattering in Ultrathin-Body SOI MOSFETs,'' 
\emph{IEEE Trans. Electron Devices}, vol.~54, no.~9, pp.~2191-2203, Sept. 2007,
doi: 10.1109/TED.2007.902712.

\bibitem{Jin2007}  
S.~Jin, M.~V.~Fischetti, and T-w.~Tang,``Modeling of electron mobility in gated silicon nanowires at room temperature: Surface Roughness scattering, dielectric screening, and band nonparabolicity,'' \emph{J. Appl. Phys.}, vol.~102, no.~8, 083715, Oct 2007, 
doi: 10.1063/1.2802586.

\bibitem{Jungemann1993}  
C.~Jungemann, A.~Emunds, and W.~L.~Engl, ``Simulation of linear and nonlinear electron transport in homogeneous silicon inversion layers,'' \emph{Solid State Electron.}, vol.~36, no.~11, pp. 1529--1540, Nov. 1993, 
doi: 10.1016/0038-1101(93)90024-K.

\bibitem{Pirovano1999}  
A.~Pirovano, A.~L.~Lacaita, G.~Zandler and R.~Oberhuber, ``Explaining the dependences of electron and hole mobilities in Si MOSFET's inversion layer,'' 
 \emph{IEDM Tech. Dig.}, Dec. 1999, pp. 527--530, 
doi: 10.1109/IEDM.1999.824208.

\bibitem{Gamiz1999}  
F.~G\'{a}miz, J.~B.~Rold\'{a}n, J.~A.~L\'{o}pez-Villanueva, P.~Cartujo-Cassinello, J.~E.~Carceller;,
``Surface roughness at the Si/SiO$_2$ interfaces in fully depleted silicon-on-insulator inversion layers,''
\emph{J. Appl. Phys.}, vol.~86, no.~12, pp.~6854--6863, Dec. 1999, 
doi: 10.1063/1.371763.

\bibitem{Esseni2004} 
D.~Esseni, ``On the modeling of surface roughness limited mobility in SOI MOSFETs and its correlation to the transistor effective field'' \emph{IEEE Trans. Electron Devices}, vol.~51, no.~3,  pp.~394-401, March 2004,
doi: 10.1109/TED.2003.819256.

\bibitem{Low2005}  
T.~Low, M.-F.~Li, G.~Samudra,Y.-C.~Yeo, C.~Zhu, and A.~Chin,“Modeling study of the impact of surface roughness on silicon and germanium UTB MOSFETs,” \emph{IEEE Trans. Electron Devices}, vol.~52, no.~11, pp.~2430--2439, Nov. 2005,
doi: 10.1109/TED.2005.857188.

\bibitem{Ishihara2006} 
T.~Ishihara, K.~Uchida, J.~Koga, and S.-I.~Takagi, ``Unified roughness scattering model incorporating scattering component induced by thickness fluctuations in silicon-on-insulator metal-oxide-semiconductor field-effect transistors,'' \emph{Jpn. J. Appl. Phys.}, vol.~45,
no.~4B, pp.~3125--3132, Apr. 2006, 
doi: 10.1143/JJAP.45.3125.

\bibitem{Poljak2012}
M.~Poljak, V.~Jovanovic, D.~Grgec, and T.~Suligoj, ``Assessment of Electron Mobility in Ultrathin-Body InGaAs-on-Insulator MOSFETs Using Physics-Based Modeling,''
\emph{IEEE Trans. Electron Devices}, vol.~59, no.~6, pp.~1636--1643, June 2012,
doi: 10.1109/TED.2012.2189217

%
{\color{black}
\bibitem{Kannan2017}
G.~Kannan and D.~Vasileska, ``The impact of surface-roughness scattering on the low-field electron mobility in nano-scale Si MOSFETs,” \emph{J. Appl. Phys.}, vol.~122, no.~24, 114303, Sept. 2017, 
doi: 10.1063/1.5003253.
}%

\bibitem{Lizzit2014}
D.~Lizzit, D.~Esseni, P.~Palestri, and L.~Selmi, ``A new formulation for Surface Roughness limited mobility in bulk and ultra-thin-body metal-oxide-semiconductor transistors,'' 
\emph{J. Appl. Phys.}, vol.~116, no.~22, 223702, Dec. 2014, 
doi: 10.1063/1.4903768.

\bibitem{Badami2016}
O.~Badami, E.~Caruso, D.~Lizzit, P.~Osgnach, D.~ Esseni, P.~Palestri, ``An improved surface roughness scattering model for bulk, thin-body, and quantum-well MOSFETs,'' \emph{IEEE Trans. Electron Devices}, vol.~63, no.~6, pp.~2306--2312, Jun. 2016,
doi: 10.1109/TED.2016.2554613.

%
{\color{black}
\bibitem{Lizzit2017}
D.~Lizzit, D.~Esseni, P.~Palestri, and L.~Selmi, ``Improved surface-roughness scattering and mobility models for multi-gate FETs with arbitrary cross-section and biasing scheme,” \emph{J. Appl. Phys.}, vol.~121, no.~24, 245301, Jun. 2017, 
doi: 10.1063/1.4986644.
}%
  
\bibitem{Sano2021PRE}
N.~Sano, ``Quantum kinetic equation for the Wigner function and reduction to the Boltzmann transport equation under discrete impurities,'' 
\emph{Phys. Rev. E}, vol.~104, no.~1, 014141, July 2021, 
doi: 10.1103/PhysRevE.104.014141.

\bibitem{Jacoboni2010book}
C.~Jacoboni, \emph{Theory of Electron Transport in Semiconductors: A Pathway
  from Elementary Physics to Nonequilibrium Green Functions}.
  New York, USA: Springer, 2010. 

\bibitem{Esseni2011}
D.~Esseni, P.~Palestri, and L.~Selmi, \emph{Nanoscale MOS Transistors: Semi-Classical Transport and Applications}, Cambridge, U.K.: Cambridge Univ. Press, 2011.

\bibitem{Doniach1998}
S.~Doniach, S. and E.H.~Sondheimer, \emph{Green's Functions for Solid State Physicists}.
  London, UK: Imperial College Press, 1998. 
  
\bibitem{Sano2025PRB}
N.~Sano, ``Nonequilibrium Green's function formalism applicable to discrete impurities in semiconductor nanostructures,'' 
\emph{Phys. Rev. B}, vol.~111, no.~12, 125413, March 2025, 
doi: 10.1103/PhysRevB.111.125413.

%
{\color{black}
\bibitem{Prange1968} 
R.E.~Prange and T.-W.~Nee, ``Quantum spectroscopy of the low-field oscillations in the surface impedeace,'' 
\emph{Phys. Rev.}, vol.~168, no.~3, pp.~779--786, April 1968, 
doi: 10.1103/PhysRev.168.779.
}%

\bibitem{Sano2025TED}
N.~Sano, ``Fundamental aspects of semiconductor device modeling associated with discrete impurities: Nonequilibrium Green's Function Scheme,'' \emph{IEEE Trans. Electron Devices}, vol.~72, no.~1, pp. 24--30, Jan. 2025, 
doi: 10.1109/TED.2024.3499940.

%

\bibitem{Mahan2000}
G.~D.~Mahan,  \emph{Many-Particle Physics, 3rd ed.}. New York, USA: Springer, 2000.

\bibitem{Datta2005}
S.~Datta, \emph{Quantum Transport: Atom to Transistor}.
  Cambridge, UK: Cambridge University Press, 2005. 

\bibitem{Haug2008}
H.~Haug and A.-P.~Jauho, \emph{Quantum Kinetics in Transport and Optics of Semiconductors}.
  New York, USA: Springer Verlag, 2008. 

 



\end{thebibliography}
%

\end{document}